\documentstyle[12pt]{article}

\newcommand{\hh}{\rule[-5mm]{0mm}{10mm}}

\newcommand{\bz}{{\hbox{\bf Z}}}
\newcommand{\br}{{\hbox{\bf R}}}
\newcommand{\bc}{{\hbox{\bf C}}}
\newcommand{\bq}{{\hbox{\bf Q}}}

\newcommand \ket[1]{\left\vert\, {#1} \, \right>}

\newcommand \qint[1]{\left[ {#1} \right]}
\newcommand \diff[2]{{~}_{\scriptstyle {#1}}
\displaystyle \partial_{\scriptstyle {#2}} \,}
\newcommand \fra[2]{\displaystyle
{\frac{\textstyle {#1}}{\textstyle {#2}}}}
\newcommand \fr[2]{\textstyle {\textstyle
{\frac{\scriptstyle {#1}}{\scriptstyle {#2}}}}}
\newcommand {\gl}{\Bigl(}
\newcommand {\gr}{\Bigr)}

\begin{document}
\begin{flushright}
RIMS-920 \\
YITP/K-1017 \\
May 1993
\end{flushright}
\vspace{24pt}
\begin{center}
\begin{large}
{\bf Free Boson Representation of $U_{q}(\widehat{sl}_3)$}
\end{large}

\vspace{36pt}
Hidetoshi Awata\footnote{Fellow of Soryushi Shogakukai},
Satoru Odake${}^1$ \\
\vspace{6pt}
{\it Yukawa Institute for Theoretical Physics\\
Kyoto University, Kyoto 606-01, Japan}\\
\vspace{6pt}
and\\
\vspace{6pt}
Jun'ichi Shiraishi\footnote{On leave from Department of Physics,
University of Tokyo, Tokyo 113, Japan}\\
\vspace{6pt}
{\it Research Institute for Mathematical Sciences\\
Kyoto University, Kyoto 606-01, Japan}
\vspace{78pt}

\underline{ABSTRACT}
\end{center}
\vspace{1.5cm}

A representation of the quantum
affine algebra $U_{q}(\widehat{sl}_3)$ of an arbitrary
level $k$ is constructed in the Fock module of eight boson fields.
This realization reduces the Wakimoto representation in the
$q \rightarrow 1$ limit.
The analogues of the screening
currents are also obtained. They commute with the action of
$U_{q}(\widehat{sl}_3)$ modulo total differences of some fields.
\vfill
\newpage

{\bf 1. Introduction} \qquad Recently the anti-ferroelectric spin-1/2
XXZ-Hamiltonian was exactly diagonalized \cite{DFJMN} by
using the technique of $q$-vertex operators \cite{FR}.
Further, an integral
formula for correlation functions of local operators was found
in the case of spin 1/2 \cite{JMMN} with the help of a boson
representation of $U_{q}(\widehat{sl}_2)$ of level $1$ \cite{FJ}.
The technique of vertex operators was also been applied to the case
of higher-spin XXZ models\cite{IIJMNT}. Construction of the free boson
representation of $U_q(\widehat{sl}_2)$ of higher levels is crucial
to obtain the integral formula for the correlation functions in this
case.
In fact the bosonization of $U_{q}(\widehat{sl}_2)$ of an arbitrary
level was given \cite{Shi}\cite{MaAbKi} using an analogue of the
currents defined by the Drinfeld
realization \cite{Dr} of $U_{q}(\widehat{sl}_2)$.
The Drinfeld's generators are expressed in terms of three boson fields.

To study the higher rank version of the XXZ model, that is,
the vertex model associated with
the $R$ matrix of $U_q(\widehat{sl}_n)$,
we are interested in a bosonization of
$U_{q}(\widehat{sl}_n)$. In this article we construct a
bosonization of the quantum affine algebra $U_{q}(\widehat{sl}_3)$
as the first step toward this direction.
In the $q \rightarrow 1$ limit,
this new representation reduces to the Wakimoto
representation with bosonized $\beta - \gamma$ system
\cite{Wa}\cite{FMS}\cite{FFIKKS}.
Further, this Wakimoto representation has a
connection with the difference operator representation of $U_q(sl_3)$
obtained in \cite{ANO}.

We obtain two analogues of
the screening currents in terms of boson fields. These operators have
the property that they commute with the currents modulo total
differences of some operators.
Hence a suitable Jackson integral of the screening currents should
commute exactly with the currents.
A Jackson integral formula for the solution to
$q$-deformed Knizhnik-Zamolodchikov equation \cite{FR}
was found by
Matsuo \cite{Ma}
and Reshetikhin \cite{Re}. We think that there is a deep connection between
the existence of our screening currents and these Jackson integral formulas.
\vspace{1cm}

{\bf 2.  Free boson fields $a^i, b^{\mu},$
and $c^{\mu}$ $(i=1,2,$ $\mu =1,2,3)$} \qquad
In this article we consider bosonization of the Drinfeld realization of
$U_{q}(\widehat{sl}_3)$.
We construct Drinfeld's generators in terms of eight free boson fields.
Hereafter let $q$ be a generic complex number
such that
$|q|<1$. We will frequently use the following standard notation:
$$
\qint{m} = \fra{q^{m}-q^{-m}}{q-q^{-1}},
$$
for $m \in \bz$.

Let $k$ be a complex number.
Let $\{a^i_{n},b^{\mu}_{n},c^{\mu}_{n},Q^i_{a},Q^{\mu}_{b},Q^{\mu}_{c}|
n \in \bz \, ,i=1,2, \, \mu=1,2,3 \}$
be the set of operators satisfying the following commutation relations:
\begin{eqnarray}
&& \qint{a^i_{n},a^j_{m}} = \delta_{n+m,0}
     \fra{\qint{a_{ij}n}\qint{(k+3)n}}{n}, \qquad
   \qint{a^i_{0},Q^j_{a}}=a_{ij}(k+3),  \nonumber \\
&& \qint{b^{\mu}_{n},b^{\nu}_{m}} = - \delta_{\mu,\nu} \delta_{n+m,0}
     \fra{\qint{n}\qint{n}}{n}, \qquad
 \qint{b^{\mu}_{0},Q^{\nu}_{b}}=-\delta_{\mu,\nu}, \\
&& \qint{c^{\mu}_{n},c^{\nu}_{m}} = \delta_{\mu,\nu} \delta_{n+m,0}
     \fra{\qint{n}\qint{n}}{n}, \qquad
 \qint{c^{\mu}_{0},Q^{\nu}_{c}}= \delta_{\mu,\nu}, \nonumber
\end{eqnarray}
where $\left(a_{ij}\right)_{i,j=1}^2$ is the Cartan
matrix of of type $A_2$, i.e.
$a_{11}=a_{22}=2, a_{12}=a_{21}=-1$. The remaining
commutators vanish.

Let us introduce eight free boson fields $a^i, b^{\mu},$ and $c^{\mu}$
$(i=1,2, \mu=1,2,3)$ carrying parameters
$M,N \in \bz_{>0}$, $\alpha \in \br$.
Define $a^i\gl M;N|z;\alpha\gr$ $(i=1,2)$ by
\begin{equation}
a^i\gl M;N|z;\alpha\gr
= - \displaystyle \sum_{n \neq 0}
\frac{\textstyle \qint{Mn} a^i_{n}}
{\textstyle \qint{Nn}\qint{n}} z^{-n}q^{|n|\alpha}
+ \frac{\textstyle Ma^i_{0}}{\textstyle N} \log z
+ \frac{\textstyle MQ^i_{a}}{\textstyle N}.
\end{equation}
Note that this definition is slightly different from the one given
in \cite{Shi}.
We define
$b^{\mu}\gl M;N|z;\alpha\gr$, $c^{\mu}\gl M;N|z;\alpha\gr$ $(\mu=1,2,3)$
in the same way.
In the case $M=N$ we write
$$
a^i\gl z;\alpha\gr = a^i\gl M;M|z;\alpha\gr ,
$$
and likewise for $b^{\mu}\gl z;\alpha\gr, c^{\mu}\gl z;\alpha\gr$.
Furthermore we introduce $a^i_{\pm}\gl M;N|z\gr$ $(i=1,2)$ as
\begin{equation}
a^i_{\pm}\gl M;N|z\gr
= \pm \Biggl( (q-q^{-1})\displaystyle
\sum_{n {>\atop<} 0}
\fra{\qint{Mn}a^i_n}{\qint{Nn}}z^{-n}
+ \fra{M a^i_{0}}{\textstyle N} \log q \Biggr).
\end{equation}
The fields $b^{\mu}_{\pm}\gl M;N|z\gr,c^{\mu}_{\pm}\gl M;N|z\gr$ $(\mu=1,2,3)$
are defined in the same way.
We note that $a^i_{\pm}\gl M;N|z\gr$ can be expressed using
$a^i\gl 1;N|z;\alpha\gr$ as
\begin{equation}
a^i_{\pm}\gl M;N|z\gr =
    a^i\gl 1;N|q^{\pm \alpha}z;\alpha-M\gr
  - a^i\gl 1;N|q^{\pm(\alpha-M)}z;\alpha\gr ,
\end{equation}
where $\alpha$ is any real number.
We note that the newly introduced fields
$a^i_{\pm}\gl M;N|z\gr,$ $ b^{\mu}_{\pm}\gl M;N|z\gr,$
$c^{\mu}_{\pm}\gl M;N|z\gr$
are thus not independent of the fields $a^i\gl M;N|z;\alpha\gr,$
$b^{\mu}\gl M;N|z;\alpha\gr,$ $c^{\mu}\gl M;N|z;\alpha\gr$.
In the following sections, we will find
it convenient to use this new notation.
It will also be convenient to use the following shorthand notation:
\begin{eqnarray}
&& \gl \fr{M}{N}a\gr^i\gl z;\alpha\gr \equiv a^i\gl M;N|z;\alpha\gr, \\
&& \gl \fr{M}{N}a\gr^i_{\pm}\gl z\gr \equiv a^i_{\pm}\gl M;N|z\gr, \\
&&\gl b+c\gr^{\mu}\gl z;\alpha\gr \equiv
b^{\mu}\gl z;\alpha\gr + c^{\mu}\gl z;\alpha\gr.
\end{eqnarray}

{\bf 3. $q$-difference operator and the Jackson integral} \qquad
Following \cite{Shi}, we define the
$q$-difference operator with a parameter $n \in \bz_{>0}$:
$$
\diff{n}{z} f(z) \equiv \frac{\textstyle f(q^{n}z) - f(q^{-n}z)}
{\textstyle (q-q^{-1})z}.
$$
Let $p$ be a complex number such that $|p|<1$ and $s \in \bc^{\times}$.
We define the Jackson integral by
$$
\int^{s\infty}_{0} f(t) d_{p}t = s(1-p) \displaystyle
\sum^{\infty}_{m=-\infty} f(sp^{m})p^{m} ,
$$
whenever it is convergent.
If the integrand $f(t)$ is a
total difference of some function $F(t)$:
$$
f(t) = \diff{n}{t} F(t),
$$
then by taking $p=q^{2n}$, we have
$$
\int^{s\infty}_{0} f(t) d_{p}t = 0.
$$

{\bf 4. Fock module and Wick's Theorem} \qquad
First we define the Fock module.
Let $\ket{0}$ be the vector having the following properties
$$
a^i_n\ket{0}=b^{\mu}_n\ket{0}=c^{\mu}_n\ket{0}=0
\qquad i=1,2\;\;\mu=1,2,3\;\;n\geq0.
$$
Define the vectors
$$
\ket{r_1,r_2;s_1,s_2,s_3}=\exp\Biggl\{
       \sum_{i,j=1}^{2}r_ia^{-1}_{ij}\fra{Q_a^j}{k+3} +
       \sum_{\mu=1}^{3}s_{\mu}\Bigl(Q_b^{\mu}+Q_c^{\mu}\Bigr)
                    \Biggr\}\ket{0},
$$
where $a^{-1}_{ij}$ is the inverse of the Cartan matrix $a_{ij}$, and
$r_i,s_{\mu} \in \bz$ $(i=1,2,\;\mu=1,2,3)$.
Let $F$ be a free $\bq(q)$ module generated by
$\{a^i_{n},b^{\mu}_{n},c^{\mu}_{n}|n \in \bz_{<0},i=1,2,\mu=1,2,3 \}$.
Now we define the Fock modules $F_{r_1,r_2;s_1,s_2,s_3}$ by
$$
F_{r_1,r_2;s_1,s_2,s_3}=F\ket{r_1,r_2;s_1,s_2,s_3}.
$$
Further we write the total Fock module ${\cal F}$ as
$$
{\cal F}=\bigoplus_{r_1,r_2,s_1,s_2,s_3\in\bz}F_{r_1,r_2;s_1,s_2,s_3}.
$$

We regard
$\{a^i_{n},b^{\mu}_{n},c^{\mu}_{n}|n \in \bz_{\geq 0},i=1,2,\mu=1,2,3 \}$
as the set of annihilation operators,
and $\{a^i_{n},$ $b^{\mu}_{n},c^{\mu}_{n},Q^i_{a},Q^{\mu}_{b},Q^{\mu}_{c}|
n \in \bz_{< 0},i=1,2,\mu=1,2,3 \}$ that of creation operators.
We denote by $: \cdots :$ the corresponding normal ordering of operators.
For example,
$$
:\exp \Biggl\{ b^{\mu}\gl z;\alpha\gr \Biggr\}: =
\exp \left\{- \displaystyle \sum_{n < 0}
\fra{b^{\mu}_{n}}{\qint{n}} z^{-n} q^{|n|\ \alpha} \right\}
\exp \left\{- \displaystyle \sum_{n > 0}
\fra{b^{\mu}_{n}}{\qint{n}} z^{-n} q^{|n|\ \alpha} \right\}
e^{Q^{\mu}_{b}} z^{b^{\mu}_{0}} .
$$
Note that such normal ordered operators are not
well defined by themselves, because they have no meaning as a
formal power series. If, however, we regard these operators as ones acting on
the Fock module ${\cal F}$ , they have a well defined meaning.

The propagators of the boson fields $a^i$
read as follows:
\begin{equation}
\begin{array}{rl}
&\left< a^i\gl M;N|z;\alpha\gr a^j\gl M';N'|w;\beta\gr \right> \hh \\
=& - \displaystyle \sum_{n>0} \fra{\qint{Mn} \qint{M'n}
\qint{a^i_{n} , a^j_{-n}}}{\qint{Nn} \qint{N'n} \qint{n} \qint{n}}
\left( \fra{w}{z} \right)^{n} q^{(\alpha + \beta)n} +
\fra{MM'\qint{a^i_{0} , Q^j_{a}}}{NN'}\log z \hh \\
=& - \displaystyle \sum_{n>0} \fra{\qint{Mn} \qint{M'n}
\qint{a_{ij} n} \qint{(k+3)n}}{\qint{Nn} \qint{N'n} \qint{n} \qint{n}}
\left( \fra{w}{z} \right)^{n} q^{( \alpha + \beta)n} +
 \fra{MM'a_{ij}(k+3)}{NN'}\log z . \hh
\end{array}
\end{equation}
The formal power series in $w/z$ is convergent if $|w/z|<\hskip-2pt<1$.
We introduce the propagators for boson fields $b^{\mu}$
and $c^{\mu}$ in the same manner.
One can rewrite
them simply by using the logarithm.
For example,
\begin{equation}
\begin{array}{rl}
\left< b^{\mu}\gl z;\alpha\gr b^{\nu}\gl w;\beta\gr \right> =&
- \delta_{\mu,\nu}
\log (z-q^{ \alpha+\beta}w) \; , \;\;\;  |z| > |q^{\alpha+\beta}w| .
\end{array}
\end{equation}
Using these propagators, we obtain Wick's Theorem in the following form:

{\bf Proposition 1} ({\it Wick's Theorem})
\begin{eqnarray*}
&&:\exp \left\{ a^i\gl M;N|z;\alpha\gr \right\}:
:\exp \left\{ a^j\gl M';N'|w;\beta\gr \right\}: \\
=&&
\exp \left\{ \left< a^i\gl M;N|z;\alpha\gr
a^j\gl M';N'|w;\beta\gr \right> \right\}
:\exp \left\{ a^i\gl M;N|z;\alpha\gr +
a^j\gl M';N'|w;\beta\gr \right\}: .
\end{eqnarray*}
{\it There are similar formulas for $b^{\mu}$ and $c^{\mu}$.}

{\bf 5. Current algebra} \qquad
In this section we construct the $U_q(\widehat{sl}_3)$ currents
$J^{\pm}_i(z)$, $\psi_i(z)$ and $\varphi_i(z)$ $(i=1,2)$.
Let us define
the
fields $J_i^{\pm}(z)$
as follows:
\begin{eqnarray*}
J^+_1(z) &=& -:\Biggl[ \diff{1}{z} \exp\Biggl\{-c^1\gl z;0\gr \Biggr\} \Biggr]
\exp\Biggl\{-b^1\gl z;1\gr \Biggr\}: ,  \\
J^+_2(z) &=& -:\Biggl[ \diff{1}{z} \exp\Biggl\{-c^2\gl qz;0\gr \Biggr\} \Biggr]
\exp\Biggl\{-b^2\gl qz;1\gr \Biggr\}
\exp\Biggl\{b^3_+\gl z\gr -b^1_+\gl qz\gr \Biggr\}: \\
         && -:\Biggl[ \diff{1}{z} \exp\Biggl\{-c^3\gl z;0\gr \Biggr\} \Biggr]
\exp\Biggl\{-b^3\gl z;1\gr \Biggr\}
\exp\Biggl\{\gl b+c\gr^1\gl z;0\gr \Biggr\}:  ,
\end{eqnarray*}
\begin{eqnarray}
 J^-_1(z)
&=& :\Biggl[ \diff{k+3}{z}
  \exp\Biggl\{\gl\fr{1}{k+3}a\gr^1 \gl z;-\fr{k+3}{2}\gr
            +\gl\fr{k+2}{k+3}b\gr^1 \gl z;-1\gr
            +\gl\fr{k+1}{k+3}c\gr^1 \gl z;0\gr  \nonumber\\
&&      \qquad\qquad\qquad      +\gl\fr{1}{k+3}b\gr^3 \gl z;-k-3\gr
            -\gl\fr{1}{k+3}b\gr^2 \gl z;-k-2\gr \Biggr\}\Biggr]  \nonumber\\
  &&  \times \exp\Biggl\{ -\gl \fr{1}{k+3}a\gr^1 \gl z;\fr{k+3}{2}\gr
            +\gl\fr{1}{k+3}b\gr^1 \gl z;-1\gr
            +\gl\fr{2}{k+3}c\gr^1 \gl z;0\gr  \\
&&       \qquad\qquad\qquad     -\gl\fr{1}{k+3}b\gr^3 \gl z;0\gr
            +\gl\fr{1}{k+3}b\gr^2 \gl z;1\gr \Biggr\}: \nonumber\\
&+&  :\Biggl[ \diff{1}{z} \exp\Biggl\{-c^2\gl q^{k+2}z;0\gr\Biggr\} \Biggr]
       \exp\Biggl\{-b^2\gl q^{k+2}z;1\gr\Biggr\}
       \exp\Biggl\{\gl b+c\gr^3\gl q^{k+2}z;0\gr\Biggr\} \nonumber\\
  && \times \exp\Biggl\{ a^1_+\gl q^{\fr{k+3}{2}}z\gr
              + b^3_+\gl q^{k+3}z\gr-b^2_+\gl q^{k+2}z\gr \Biggr\}:,\nonumber\\
J^-_2(z)
&=&  :\Biggl[ \diff{k+3}{z}
  \exp\Biggl\{\gl\fr{1}{k+3}a\gr^2 \gl z;-\fr{k+3}{2}\gr
            +b^2 \gl z;-1\gr
            +\gl\fr{k+2}{k+3}c\gr^2 \gl z;0\gr \Biggr\} \Biggr] \nonumber\\
  &&  \times \exp\Biggl\{ -\gl\fr{1}{k+3}a\gr^2 \gl z;\fr{k+3}{2}\gr
            +\gl \fr{1}{k+3}c\gr^2\gl z;0\gr \Biggr\}:   \nonumber\\
&-&  :\Biggl[ \diff{1}{z} \exp\Biggl\{-c^1\gl q^{-k-1}z;0\gr\Biggr\} \Biggr]
       \exp\Biggl\{-b^1\gl q^{-k-1}z;-1\gr\Biggr\}
       \exp\Biggl\{\gl b+c\gr^3\gl q^{-k-1}z;0\gr\Biggr\} \nonumber\\
  && \times \exp\Biggl\{ a^2_-\gl q^{-\fr{k+3}{2}}z\gr
                    +\gl \fr{2}{1}b\gr^2_-\gl q^{-k-2}z\gr \Biggr\}: .\nonumber
\end{eqnarray}
Define further the
fields $\psi_i(z), \varphi_i(z)$ as
\begin{eqnarray}
\psi_1(z) &=&
\exp\left\{a^1_+\gl q^{\fr{3}{2}}z\gr
        +\gl\fr{2}{1}b\gr^1_+\gl q^{\fr{k}{2}+1}z\gr
        -b^2_+\gl q^{\fr{k}{2}+2}z\gr
        +b^3_+\gl q^{\fr{k}{2}+3}z\gr \right\} ,\nonumber\\
\varphi_1(z) &=&
\exp\left\{a^1_-\gl q^{-\fr{3}{2}}z\gr
        +\gl\fr{2}{1}b\gr^1_-\gl q^{-\fr{k}{2}-1}z\gr
        -b^2_-\gl q^{-\fr{k}{2}-2}z\gr
        +b^3_-\gl q^{-\fr{k}{2}-3}z\gr \right\} , \\
\psi_2(z) &=&
\exp\left\{a^2_+\gl q^{\fr{3}{2}}z\gr
        -b^1_+\gl q^{\fr{k}{2}+1}z\gr
        +\gl\fr{2}{1}b\gr^2_+\gl q^{\fr{k}{2}+2}z\gr
        +b^3_+\gl q^{\fr{k}{2}}z\gr \right\} ,\nonumber\\
\varphi_2(z) &=&
\exp\left\{a^2_-\gl q^{-\fr{3}{2}}z\gr
        -b^1_-\gl q^{-\fr{k}{2}-1}z\gr
        +\gl\fr{2}{1}b\gr^2_-\gl q^{-\fr{k}{2}-2}z\gr
        +b^3_-\gl q^{-\fr{k}{2}}z\gr\right\} .\nonumber
\end{eqnarray}

The formulas (10) are not so useful for OPE calculation, because
they contain difference operators, and bosons in these formulas are
somewhat complicated.
By the definition of the boson fields $a^i,b^{\mu},c^{\mu}$,
$a^i_{\pm},b^{\mu}_{\pm},c^{\mu}_{\pm}$,
and the
$q$-difference operator,
we can recast the fields $J^{\pm}_i(z)$ as
\begin{eqnarray*}
J_1^+(z) = \fra{-1}{(q-q^{-1})z} &:\Biggl(
            & \exp\Biggl\{b^1_+\gl z\gr - \gl b+c\gr^1\gl qz;0\gr\Biggr\}  \\
            &-&\exp\Biggl\{b^1_-\gl z\gr
              - \gl b+c\gr^1\gl q^{-1}z;0\gr\Biggr\} \Biggr): , \\
J_2^+(z) = \fra{-1}{(q-q^{-1})z} &:\Biggl(
            & \exp\Biggl\{-b^1_+\gl qz\gr + b^2_+\gl qz\gr + b^3_+\gl z\gr
             - \gl b+c\gr^2\gl q^2z;0\gr\Biggr\} \\
            &-&\exp\Biggl\{-b^1_+\gl qz\gr + b^2_-\gl qz\gr + b^3_+\gl z\gr
             - \gl b+c\gr^2\gl z;0\gr\Biggr\} \\
            &+&\exp\Biggl\{b^3_+\gl z\gr + \gl b+c\gr^1\gl z;0\gr
              - \gl b+c\gr^3\gl qz;0\gr\Biggr\} \\
            &-&\exp\Biggl\{b^3_-\gl z\gr + \gl b+c\gr^1\gl z;0\gr
               - \gl b+c\gr^3\gl q^{-1}z;0\gr\Biggr\}
             \Biggr): ,
\end{eqnarray*}
\begin{eqnarray*}
J_1^-(z) = \fra{1}{(q-q^{-1})z} &:\Biggl(
            & \exp\Biggl\{a^1_+\gl q^{\fr{k+3}{2}}z\gr
                          + b^1_+\gl q^{k+2}z\gr
                          - b^2_+\gl q^{k+2}z\gr
                          + b^3_+\gl q^{k+3}z\gr  \\
               && \qquad\qquad + \gl b+c\gr^1\gl q^{k+1}z;0\gr\Biggr\}  \\
          &-& \exp\Biggl\{a^1_-\gl q^{-\fr{k+3}{2}}z\gr
                          + b^1_-\gl q^{-k-2}z\gr
                          - b^2_-\gl q^{-k-2}z\gr
                          + b^3_-\gl q^{-k-3}z\gr \\
            && \qquad\qquad   + \gl b+c\gr^1\gl q^{-k-1}z;0\gr\Biggr\}  \\
          &+& \exp\Biggl\{a^1_+\gl q^{\fr{k+3}{2}}z\gr
                          + b^3_+\gl q^{k+3}z\gr  \\
             && \qquad\qquad      - \gl b+c\gr^2\gl q^{k+3}z;0\gr
                          + \gl b+c\gr^3\gl q^{k+2}z;0\gr\Biggr\}  \\
          &- & \exp\Biggl\{a^1_+\gl q^{\fr{k+3}{2}}z\gr
                          - b^2_+\gl q^{k+2}z\gr
                          + b^2_-\gl q^{k+2}z\gr
                          + b^3_+\gl q^{k+3}z\gr \\
            && \qquad\qquad   - \gl b+c\gr^2\gl q^{k+1}z;0\gr
                          + \gl b+c\gr^3\gl q^{k+2}z;0\gr\Biggr\}  \Biggr): ,\\
J_2^-(z) = \fra{1}{(q-q^{-1})z} &:\Biggl(
            & \exp\Biggl\{a^2_+\gl q^{\fr{k+3}{2}}z\gr
                          + b^2_+\gl q^{k+3}z\gr
                          + \gl b+c\gr^2\gl q^{k+2}z;0\gr\Biggr\}  \\
          &- & \exp\Biggl\{a^2_-\gl q^{-\fr{k+3}{2}}z\gr
                          + b^2_-\gl q^{-k-3}z\gr
                          + \gl b+c\gr^2\gl q^{-k-2}z;0\gr\Biggr\}  \\
          &+ & \exp\Biggl\{a^2_-\gl q^{-\fr{k+3}{2}}z\gr
                          - b^1_+\gl q^{-k-1}z\gr
                          + \gl \fr{2}{1}b\gr^2_-\gl q^{-k-2}z\gr \\
              && \qquad\qquad     - \gl b+c\gr^1\gl q^{-k-2}z;0\gr
                          + \gl b+c\gr^3\gl q^{-k-1}z;0\gr\Biggr\}  \\
          &- & \exp\Biggl\{a^2_-\gl q^{-\fr{k+3}{2}}z\gr
                          - b^1_-\gl q^{-k-1}z\gr
                          + \gl \fr{2}{1}b\gr^2_-\gl q^{-k-2}z\gr \\
                && \qquad\qquad  - \gl b+c\gr^1\gl q^{-k}z;0\gr
                          + \gl b+c\gr^3\gl q^{-k-1}z;0\gr\Biggr\}  \Biggr): .
\end{eqnarray*}

Our main purpose is to state the OPE algebra of these currents.
To this end, let
us introduce the function $g_{ij}(z)$ as the following
formal power series
$$
g_{ij}(z)=\Bigl(q^{-a_{ij}}-z\Bigr)\times
\sum_{n\geq0}\left(q^{-a_{ij}}z\right)^n,
$$
and its inverse
$$
g_{ij}(z)^{-1}=\Bigl(q^{a_{ij}}-z\Bigr)\times
\sum_{n\geq0}\left(q^{a_{ij}}z\right)^n.
$$
Further define $\delta(z)$ by
\begin{equation}
\delta(z)=\sum_{n\in\bz}z^n.
\end{equation}
Using Wick's Theorem, we get the following formulas.

{\bf Proposition 2}
{\it Let $J^{\pm}_i(z), \psi_i(z)$ and $\varphi_i(z)$ $(i=1,2)$ be
the fields defined as above, and let $\gamma = q^{k}$, then
the following relations hold in the sense of a formal power series:}
\begin{eqnarray*}
&& \qint{\varphi_i (z), \varphi_j (w)} = 0, \\
&& \qint{\psi_i (z), \psi_j (w)} = 0, \\
&& \varphi_i (z) \psi_j (w) =g_{ij}(zw^{-1}\gamma^{-1})
g_{ij}(zw^{-1}\gamma)^{-1}
\psi_j (w) \varphi_i (z), \\
&& \varphi_i (z) J^{\pm}_j(w) =g_{ij}(zw^{-1}\gamma^{\mp1/2})^{\pm1}
J^{\pm}_j(w) \varphi_i(z), \\
&& \psi_i (z) J^{\pm}_j(w) =g_{ij}(z^{-1}w\gamma^{\mp1/2})^{\mp1}
J^{\pm}_j(w) \psi_i(z), \\
&& \qint{J^+_i(z), J^-_i(w)} = \fra{\delta_{ij}}{(q-q^{-1})zw}
\left(\delta(zw^{-1}\gamma^{-1})\psi_i(w\gamma^{1/2}) -
\delta(zw^{-1}\gamma)\varphi_i(w\gamma^{-1/2})
\right), \\
&& (z-wq^{\pm a_{ij}})J^{\pm}_i(z)J^{\pm}_j(w) =
(zq^{\pm a_{ij}}-w)J^{\pm}_j(w)J^{\pm}_i(z) , \\
&&\Bigl\{
J^{\pm}_i(z_1)J^{\pm}_i(z_2)J^{\pm}_j(w) -(q+q^{-1})
J^{\pm}_i(z_1)J^{\pm}_j(w)J^{\pm}_i(z_2) +
J^{\pm}_j(w)J^{\pm}_i(z_1)J^{\pm}_i(z_2)
\Bigr\} \\
&&+ \Bigl\{z_1 \leftrightarrow z_2 \Bigr\}
= 0 \qquad \mbox{for } a_{ij}=-1.
\end{eqnarray*}

Before we consider the mode expansions of the fields
$J^{\pm}_i(z), \psi_i(z),$ and $\varphi_i(z)$, let us introduce
the operators $J^3_{in}$ ($n \in \bz_{\neq 0}$) and $K^{\pm1}_i$ as
\begin{eqnarray*}
&& J^3_{1n} =  a^1_n q^{-\fr{3}{2}|n|}
 +   \frac{\qint{2n}}{\qint{n}} b^1_n q^{-\left(\fr{k}{2}+1\right)|n|}
 -   b^2_n q^{-\left(\fr{k}{2}+2\right)|n|}
 +   b^3_n q^{-\left(\fr{k}{2}+3\right)|n|}, \\
&& J^3_{2n} =  a^2_n q^{-\fr{3}{2}|n|}
 -   b^1_n q^{-\left(\fr{k}{2}+1\right)|n|}
 +   \frac{\qint{2n}}{\qint{n}} b^2_n q^{-\left(\fr{k}{2}+2\right)|n|}
 +   b^3_n q^{-\fr{k}{2}|n|}, \\
&& K^{\pm1}_i=
        q^{\pm \left(a^i_{0}+\sum_{j=1}^2 a_{ij}b^j_{0} + b^3_{0}\right)}.
\end{eqnarray*}
Then we can write $\psi_i(z)$ and $\varphi_i(z)$ as follows
\begin{eqnarray*}
&& \psi_i(z) = K_i \exp\Biggl\{
(q-q^{-1})\sum_{n>0} J^3_{in} z^{-n} \Biggr\}, \\
&& \varphi_i(z) = K_i^{-1} \exp\Biggl\{
-(q-q^{-1})\sum_{n<0} J^3_{in} z^{-n} \Biggr\}.
\end{eqnarray*}

Since the currents $J^{\pm}_i(z), \psi_i(z),$ and $\varphi_i(z)$ are
well defined operators acting on the Fock module ${\cal F}$,
then we can consider the following mode expansions
\begin{equation}
 \sum_{n \in \bz} J^{\pm}_{in}z^{-n-1}=J^{\pm}_i(z),~~~
 \sum_{n \in \bz} \psi_{in} z^{-n} = \psi_i(z),~~~
 \sum_{n \in \bz} \varphi_{in} z^{-n} = \varphi_i(z).
\end{equation}

Now we are ready to state our main proposition:

{\bf Proposition 3} {\it The operators
$\{J^{3}_{in}|n \in \bz_{\neq 0},i=1,2 \}, \{ J^{\pm}_{in}|
n \in \bz,i=1,2 \}$, $K_i$ $(i=1,2)$
and $\gamma^{\pm1/2}$ acting on the Fock module ${\cal F}$,
satisfy the following relations.}
\begin{eqnarray*}
&& \gamma^{\pm1/2} \in \mbox{ the center of the algebra}, \\
&& \qint{J^3_{in},J^3_{im}}
= \delta_{n+m,0} \fra{1}{n}
\qint{a_{ij}n}\fra{\gamma^{n}-\gamma^{-n}}{q-q^{-1}}, \\
&& \qint{J^3_{in},K_j} = 0, \\
&&  K_i J^{\pm}_{jn} K^{-1}_i = q^{\pm a_{ij}} J^{\pm}_{jn}, \\
&& \qint{J^3_{in},J^{\pm}_{jm}}
= \pm \fra{1}{n}\qint{a_{ij}n}\gamma^{\mp|n|/2} J^{\pm}_{jn+m}, \\
&& J^{\pm}_{in+1} J^{\pm}_{jm} - q^{\pm a_{ij}} J^{\pm}_{jm} J^{\pm}_{in+1}
= q^{\pm a_{ij}} J^{\pm}_{in} J^{\pm}_{jm+1} - J^{\pm}_{jm+1} J^{\pm}_{in}, \\
&& \qint{J^+_{in}, J^-_{jm}}
= \fra{\delta_{i,j}}{q-q^{-1}}\left(
    \gamma^{(n-m)/2}\psi_{in+m} - \gamma^{(m-n)/2}\varphi_{in+m}
                    \right), \\
&& \Bigl\{ J^{\pm}_{il}J^{\pm}_{im}J^{\pm}_{jn}
-(q+q^{-1})J^{\pm}_{il}J^{\pm}_{jn}J^{\pm}_{im}
+J^{\pm}_{jn}J^{\pm}_{il}J^{\pm}_{im}\Bigr\} \\
&& + \Bigl\{l \leftrightarrow m \Bigr\}=0 \qquad \mbox{ for  } a_{ij}=-1.
\end{eqnarray*}
These are exactly the relations of the Drinfeld realization of
$U_{q}(\widehat{sl}_3)$ for level $k$ \cite{Dr}.
Thus (10) and (11) yields the required bosonization.
Since the vectors $\ket{l_1,l_2;0,0,0}$ ($l_1,l_2\in\bz$) have the following
properties
\begin{eqnarray*}
&&K_i\ket{l_1,l_2;0,0,0}=q^{l_i}\ket{l_1,l_2;0,0,0}\qquad i=1,2,\\
&&J^{+}_{in}\ket{l_1,l_2;0,0,0}=0 \qquad i=1,2 \;\; n\geq0,\\
&&J^{-}_{im}\ket{l_1,l_2;0,0,0}=0 \qquad i=1,2 \;\; m>0,\\
&&\psi_{im}\ket{l_1,l_2;0,0,0}=0 \qquad i=1,2 \;\; m>0,
\end{eqnarray*}
we have the highest weight representations of $U_q(\widehat{sl}_3)$ in the
Fock module ${\cal F}$.
One can immediately find that this representation reduces to the
Wakimoto representation in the $q \rightarrow 1$ limit.
Note that we have a nice relation between this Wakimoto representation
and the difference operator representation of $U_q(sl_3)$ given
in ref.\cite{ANO}.

{\bf 6. screening currents} \qquad
Let us define the screening currents $S_i(z)$ $(i=1,2)$ as follows:
\begin{eqnarray}
S_1(z) &=& -:\Bigg(\Biggl[
   \diff{1}{z} \exp\Biggl\{-c^1\gl qz;0\gr\Biggr\} \Biggr]
\exp\Biggl\{-b^1\gl qz;-1\gr\Biggr\}
\exp\Biggl\{-b^3_-\gl z\gr+b^2_-\gl qz\gr\Biggl\} \nonumber\\
 && \qquad +\Biggl[ \diff{1}{z} \exp\Biggl\{-c^3\gl z;0\gr\Biggr\} \Biggr]
\exp\Biggl\{-b^3\gl z;-1\gr\Biggr\}
\exp\Biggl\{\gl b+c\gr^2(z;0)\Biggr\} \Bigg) \\
 && \times
  \exp\Biggl\{-\gl \fr{1}{k+3}a\gr^1\gl z;-\fr{k+3}{2}\gr\Biggr\}:,
\nonumber\\
S_2(z) &=& -:\Biggl[ \diff{1}{z} \exp\left\{-c^2\gl z;0\gr\right\} \Biggr]
\exp\Biggl\{-b^2\gl z;-1\gr\Biggr\}
\exp\Biggl\{-\gl \fr{1}{k+3}a\gr^2\gl z;-\fr{k+3}{2}\gr\Biggr\}:.
\nonumber
\end{eqnarray}
Then we get the following proposition and its corollary.

{\bf Proposition 4}
{\it The following commutation relations hold.}
\begin{eqnarray*}
&& \qint{\psi_i(z), S_j(w) } = 0,  \\
&& \qint{\varphi_i(z), S_j(w)}  = 0,  \\
&& \qint{J^{+}_i(z),  S_j(w)}  =  0, \\
&& \qint{J^{-}_i(z), S_j(w)} =
\fra{\delta_{ij}}{z} \diff{k+3}{w} \Biggl(\delta(w/z)
:\exp\Biggl\{ -\gl \fr{1}{k+3}a\gr^i\gl w;\fr{k+3}{2}\gr\Biggr\}: \Biggr).
\end{eqnarray*}

{\bf Corollary 5}
{\it If the Jackson integral of the screening currents}
$$
\int^{s\infty}_{0} S_i(t) d_{p}t , \;\;\; p=q^{2(k+3)}
$$
{\it are convergent, then they commute with
the action of $U_{q}(\widehat{sl}_3)$ exactly.}

Results obtained in this article can be extended to higher rank
algebra $U_q(\widehat{sl}_n)$ \cite{AOS}.
\vspace{1cm}

{\bf Acknowledgments} \qquad
The authors would like to thank T. Inami for careful reading
the manuscript and discussions.
They would also like to thank T. Eguchi, M. Jimbo,
T. Miwa, A. Nakayashiki, M. Noumi, A. W. Schnizer, V.Tarasov and
Y. Yamada for helpful discussions.

\qquad

\end{document}